\documentclass[]{article}
\usepackage{epsfig}
\usepackage{wrapfig}

\def\lsim{\raise0.3ex\hbox{$<$\kern-0.75em\raise-1.1ex\hbox{$\sim$}}}
\def\gsim{\raise0.3ex\hbox{$>$\kern-0.75em\raise-1.1ex\hbox{$\sim$}}}
\begin{document}

\begin{center}

\uppercase{\large{\bf Photoproduction of isolated photons, single hadrons and jets at
NLO}}\footnote{To appear in the proceedings of the Ringberg Workshop 
"New Trends in HERA Physics 2003", Ringberg Castle, Tegernsee, 
Germany, September 28 - October 3, 2003.}

\vskip  0.5 truecm 

\large{\bf Gudrun Heinrich}\\[.5cm]

{\em II. Institut f\"ur theoretische Physik, 
Universit\"at Hamburg,\\ 
Luruper Chaussee 149, 
22761 Hamburg, 
Germany}

\vskip  0.5 truecm 

\begin{abstract}
The photoproduction of large-$p_T$ charged hadrons 
and of prompt photons is discussed, for the inclusive case 
and with an associated jet,   
using predictions from the NLO partonic Monte Carlo program 
EPHOX. Comparisons to recent HERA data are also shown.  
\end{abstract}

\end{center}

\section{Introduction}

At HERA, the photoproduction of jets 
has been measured with high statistics 
over the past years\,\cite{Adloff:2003nr,Chekanov:2002ru,cvach}. 
Data on  and single charged hadron\,\cite{Adloff:1998vt} and prompt 
photon\,\cite{Breitweg:1999su,Chekanov:2001aq,h1} 
production are also available and have been compared 
to theoretical predictions\,\cite{Kniehl:2000hk}-\,\cite{Zembrzuski:2003nu}, 
%\,\cite{Kniehl:2000hk,Fontannaz:2002nu,Gordon:1995km,Gordon:1997yt,
%Fontannaz:2001ek,Fontannaz:2001nq,Krawczyk:2001tz,Zembrzuski:2003nu}, 
%for the inclusive case~\,\cite{} as well as for the 
%production of the $\gamma$\,+jet 
%cross section~\,\cite{}, 
where in the prompt photon case the statistics is of 
course lower as the cross sections are small. 

In photoproduction reactions, an almost real photon,  
emitted at small angle from the electron, 
interacts with a parton from the proton. 
The photon can either participate directly in the 
hard scattering ("direct photon")
or act as a  "resolved photon",  
in which case a parton stemming from the
photon takes part in the hard interaction. 
Therefore photoproduction reactions offer 
a unique opportunity to constrain the parton distributions
in the photon, especially the  
gluon distribution $g^{\gamma}(x)$. 
The latter is rather poorly constrained by  LEP $\gamma^*\gamma$
data as it enters in this reaction  only at next-to-leading order (NLO), 
whereas in $e\,p$ photoproduction it enters already at leading order. 
%through subprocesses like $g^\gamma\,q^p\to \gamma \,q$.  

On the other hand, photoproduction reactions at HERA 
also could serve to constrain 
the gluon distribution in the {\it proton}. 
To determine the latter to a better accuracy 
is of particular interest in view of the LHC with its large gluon 
luminosity, $gg\to H$ being the dominant production mode 
for a light Higgs boson. 
As the  photon represents a rather "clean" initial state, 
photoproduction reactions can be used to probe the proton 
in a way which is complementary to other experiments. 
%, provided the uncertainty from the poorly known gluon 
%in the resolved photon can be suppressed 
%for this purpose. 
%In what follows, it will be shown that appropriate cuts can 
%indeed enhance the gluon in the proton while keeping the uncertainty 
%from the photon side small. 

In what follows, the kinematic regions where gluon initiated 
subprocesses play a significant role will be identified and 
it will be argued that their relative contribution to the cross 
section can be enhanced by appropriate cuts, for the proton as 
well as for the photon case. For a more detailed study, 
the reader is referred to\,\cite{new}.

Comparisons to HERA data, especially to very recent (preliminary) 
H1 data on prompt photon plus jet production, will also be shown.

\section{Theoretical aspects of charged hadron and prompt photon production}

We will concentrate here on the two reactions
% the photoproduction of a single charged hadron in association with a jet, 
$\gamma \,p \to h^{\pm}$\,(+\,jet)\,+\,X  
and 
%the photoproduction of a prompt photon plus a jet. 
$\gamma \,p \to \gamma$\,(+\,jet)\,+\,X.
Identifying a jet in addition to the prompt photon or hadron 
allows for a more detailed study of the underlying parton dynamics, 
and in particular for the definition of observables $x_{obs}, x_{LL}$ 
which are suitable to study the parton distribution functions. 

Comparing the reaction of charged hadron production versus 
prompt photon production, one is tempted to say that 
prompt photon production is more advantageous: Due to 
photon isolation, the dependence of the 
theoretical prediction on the non-perturbative fragmentation functions 
is negligible. Further, 
as will be shown in the following, 
the NLO prediction for the prompt photon cross section
is not very sensitive to scale changes, which cannot be said for 
the $h^{\pm}$\,(+ jet) cross section\,\cite{Fontannaz:2002nu}. 
On the other hand, the $\gamma$\,+ jet cross section is orders of 
magnitude smaller than the $h^{\pm}$\,+ jet cross section, 
and the identification of the prompt photon events among the huge background 
from the decay of neutral pions is experimentally not an easy task. 

\subsection*{Photon isolation}

Prompt photons can originate 
from two mechanisms: Either they stem directly from the hard interaction, 
or they are produced by  the fragmentation of a hard parton. 
In order to distinguish them from the background stemming from the decay of 
light mesons, an {\it isolation criterion} has to be imposed:  
The amount of hadronic transverse energy
$E_T^{\rm had}$, deposited inside a cone with aperture $R$ 
centered around the photon direction 
in the rapidity and azimuthal angle plane,  
must be smaller than some value $E_T^{\rm max}$: 
\begin{equation}\label{criterion}
%\left.
\begin{array}{rcc} 
\left(  \eta - \eta^{\gamma} \right)^{2} +  \left(  \phi - \phi^{\gamma} \right)^{2}  
& \leq  & R^{2} \\
E_T^{\rm had} & \leq & E_T^{\rm max}\;.
\end{array}
%\right\} 
\end{equation}
For the numerical results we will follow the HERA conventions
$E_T^{\rm max}= 0.1\,p_T^{\gamma}$ and $R$ = 1. 

Apart from reducing the background from secondary photons, isolation also 
substantially  reduces the fragmentation component, 
%\footnote{A criterion  
%proposed in \,\cite{frixione}, in which the veto on 
%hadronic transverse energy is the more severe, the closer the corresponding
%hadron is to the photon direction, even suppresses the 
%fragmentation contribution  completely, 
%but is less straightforward to implement experimentally.},
such that the total cross 
section depends very little on the photon fragmentation functions. 

\subsection*{Factorisation and scale dependence}

%At large momentum transfers, QCD factorisation of the cross section 
%into a (non-perturbative) long-distance part and a perturbatively 
%calculable short-distance partonic cross section 
%is valid, such that 
The cross section for prompt photon 
production can be written symbolically as  
\begin{eqnarray*}
&&d\sigma^{e p \to \gamma \,X}(P_e,P_p,P_{\gamma})=\\
&&\quad\quad\sum_{a,b}\int dx_e\, d x_p\,
F_{a/e}(x_e,M)F_{b/p}(x_p,M)
\,\{d\hat\sigma^{\rm{dir}}\;+ 
\;d\hat\sigma^{\rm{frag}}\}
%\quad + {\small {\cal O}(1/Q^{\nu})}\\
\end{eqnarray*}
\begin{eqnarray*}
{d\hat\sigma^{\rm{dir}}}&=&d\hat\sigma^{ab\to \gamma\,X}
(x_a,x_b,P_{\gamma},\mu,M,M_F)\\
{d\hat\sigma^{\rm{frag}}}&=&\sum_c\int dz\,D_{\gamma/c}(z,M_F)
d\hat\sigma^{ab\to c\,X}(x_a,x_b,P_{\gamma}/z,\mu,M,M_F)\;.
\end{eqnarray*}
In the case of the production of a single hadron, the partonic 
cross section $d\hat\sigma^{\rm{dir}}$ is of course zero, 
and the fragmentation functions $D_{\gamma/c}(z,M_F)$ have 
to be replaced by the ones for hadrons, $D_{h/c}(z,M_F)$.

In photoproduction, $F_{a/e}(x_e,M)$ is a convolution
\begin{equation}
F_{a/e}(x_e,M)=\int dx^{\gamma}\int dy\,\delta(x^{\gamma}y-x_e)
\, f_{\gamma/e}(y)
\,  F_{a/\gamma}(x^{\gamma},M)\label{fae}
\end{equation}
where the spectrum of quasi-real photons emitted from the electron 
is described by the Weizs\"acker-Williams approximation
\begin{eqnarray*}
f_{\gamma/e}(y) &=& \frac{\alpha}{2\pi}\left\{\frac{1+(1-y)^2}{y}\,
{\rm Log}\frac{Q^2_{\rm max}(1-y)}{m_e^2y^2}-\frac{2(1-y)}{y}\right\}\;.
\end{eqnarray*}
%\begin{figure}[htb]
%\begin{center}
\begin{wrapfigure}[18]{r}[0.cm]{5.8cm}
\vspace*{-0.6cm}
\epsfxsize=5.8cm   %width of figure - will enlarge/reduce the figures
\epsfbox{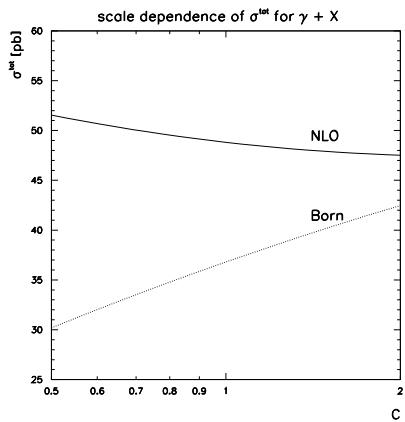}
\caption{The scales have been set to $\mu=M=M_F=C\,p_T^\gamma$ 
with $C$ varied between 0.5 and 2.\label{scaleallnobox}}
%\end{center}
\end{wrapfigure}
The function $F_{a/\gamma}(x^{\gamma},M)$ in eq.~(\ref{fae}) 
denotes the parton distribution function for a parton of type "$a$" 
in the resolved photon. 
In the case of a direct initial photon, one has 
$F_{a/\gamma}(x^{\gamma},M)=\delta_{a\gamma}\,\delta(1-x^\gamma)$. 
Therefore  
one can try to switch on/off the resolved photon by suppressing/enhancing 
large $x^\gamma$. 
 
As the perturbative series in $\alpha_s$ for the partonic cross section 
is truncated, the theoretical prediction is 
scale dependent, depending on the renormalisation scale ${\mu}$ 
as well as on the 
initial/final state factorisation scales $M$/$M_F$. 
%Of course, the scale dependence decreases the more orders 
%of the perturbative series are taken into account. 
While the leading order cross section depends very strongly on the 
scales, the NLO cross section is already much more stable, 
as can be seen for example from  Fig.\,\ref{scaleallnobox} 
for the prompt photon inclusive cross section.   

\subsection*{Observables $x^\gamma,x^p$}

In order to reconstruct the longitudinal momentum fraction 
of the parton stemming from the proton respectively the photon
from measured quantities, one can define the observables 
\begin{eqnarray}
x_{obs}^{\gamma}&=&\frac{p_T\,{\rm{e}}^{-\eta}+E_T^{jet}\,{\rm e}^{
-\eta^{jet}}}{2E^{\gamma}}\label{xobs}\\
%\quad , \quad
x_{obs}^{p}&=&\frac{p_T\,{\rm{e}}^{\eta}+E_T^{jet}\,{\rm e}^{
\eta^{jet}}}{2E^{p}}\nonumber
\end{eqnarray}
where $p_T$ is the momentum of the prompt $\gamma$ or the hadron, 
and $\eta$ its (pseudo-)rapidity. 
However, as the measurement of $E_T^{jet}$ can be a 
source of systematic errors at low $E_T$ values, 
the following variable, which does not depend on 
$E_T^{jet}$, might be more convenient: 
\begin{equation}
x_{LL}^{p,\gamma}=\frac{p_T\,({\rm e}^{\pm\eta}+{\rm e}^{
\pm\eta^{jet}})}{2E^{p,\gamma}}\label{xll}
\end{equation}
The variable $x_{LL}^{\gamma}$ also has the advantage that it 
has a smoother behaviour for $x^\gamma\to 1$.
%, as can be seen from Fig~\ref{xxllgam}. 
%\begin{figure}[ht]
%\epsfxsize=6.5cm   %width of figure - will enlarge/reduce the figures
%\epsfbox{xxll_gam.eps}
%\caption{}
%\label{xxllgam}
%\end{figure}

\section{Numerical studies}

Unless stated otherwise, the following input for the numerical 
results is used: 
The center of mass energy is $\sqrt{s}=318$\,GeV with $E_e=27.5$\,GeV and 
$E_p=920$\,GeV. 
The maximal photon virtuality is $Q^2_{\rm max}=1$\,GeV$^2$, and 
$0.2<y<0.7$.
For the parton distributions in the
proton we take the MRST01\,\cite{mrst01} parametrisation, for the photon we 
use AFG04\,\cite{Aurenche:1994in} 
distribution functions and BFG\,\cite{Bourhis:1998yu} fragmentation functions. 
For the charged hadron fragmentation functions we use BFGW\,\cite{Bourhis:2000gs} 
as default. 
We take $n_f=4$ flavours, and for $\alpha_s(\mu)$ we use an exact 
solution of the two-loop renormalisation group
equation, and not an expansion in log$(\mu/\Lambda)$. 
The default scale choice is $M=M_F=\mu=p_T$. 
Jets are defined using the $k_T$-algorithm. 

\subsection{The NLO program EPHOX}

All numerical results shown here are obtained by the NLO 
partonic  Monte Carlo program EPHOX\,\cite{ephox}, 
which allows to obtain integrated cross sections 
as well as fully differential distributions for the  
reactions considered here.
%$\gamma\,p \to \gamma +\mbox{jet} + X$ 
%$\gamma\,p \to \mbox{hadron} +\mbox{jet}+ X$, where the single particle
%inclusive case of course also is contained. 
The program contains the full NLO corrections to all four categories 
of subprocesses 
direct-direct, direct-fragmentation, resolved-direct and resolved-fragmentation.
It also  contains the quark loop box diagram 
$\gamma g\to \gamma g$ with a flag to turn it on or off. 
EPHOX can be obtained at 
{\tt http://wwwlapp.in2p3.fr/lapth/PHOX\_FAMILY/\-main.html}, 
together with a comfortable user interface and detailed documentation.

\subsection{The gluon distribution in the photon}
\begin{figure}[htb]
\begin{center}
\epsfig{file=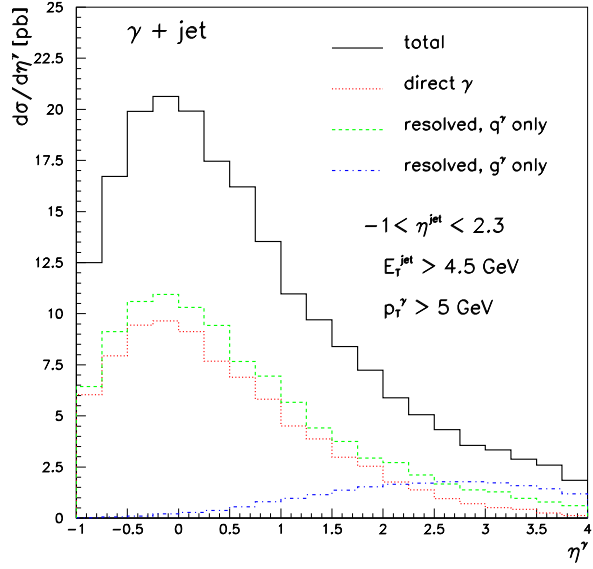,height=8.5cm}
\caption{Rapidity range where the gluon in the photon becomes important}
\label{fulleta}
\end{center}
\end{figure}

\begin{figure}[htb]
%\begin{wrapfigure}[17]{r}[0.cm]{7.5cm}
%\vspace*{-0.6cm}
\begin{center}
\epsfig{file=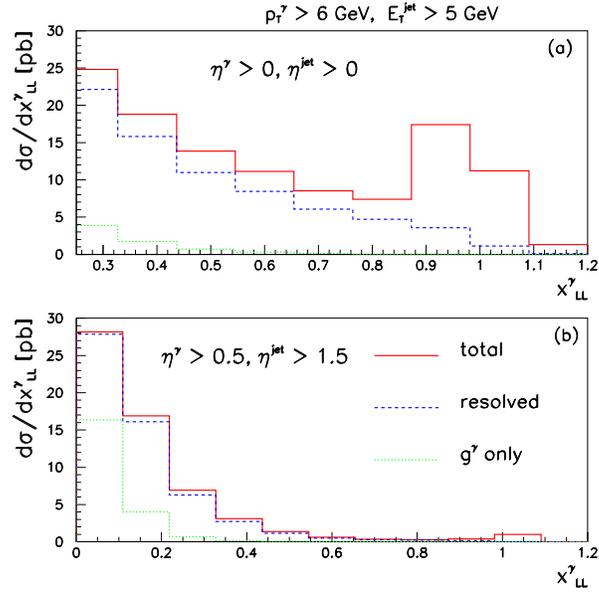,height=8.5cm}
\caption{The rapidity cuts $\eta^\gamma>0.5, \,\eta^{jet}>1.5$ 
substantially enhance the
relative contribution of the gluon in the resolved photon}
\label{xgamrapcuts}
\end{center}
\end{figure}

In order to constrain the gluon distribution $g^\gamma(x^\gamma)$
in the photon, one has to focus on a kinematic region where 
the gluon content of the resolved photon is large, i.e. on small 
values of $x^{\gamma}$. 
According to eqs.\,(\ref{xobs}) and (\ref{xll}), small $x^\gamma$ 
corresponds to 
large values of $\eta$ and $\eta^{jet}$, which means that 
the forward rapidity region is particularly interesting. 
Fig.\,\ref{fulleta} illustrates  for the 
prompt photon plus jet cross section that the gluon distribution 
$g^\gamma(x^\gamma)$
only becomes important in the very forward region $\eta^\gamma\,\gsim\, 1.5$
if the rapidity of the jet is integrated over the range $-1<\eta^{jet}<2.3$.
However, the relative importance of the gluon distribution $g^\gamma(x)$ 
can be enhanced by restricting both rapidities, 
the one of the photon {\it and } the one of the jet, 
to the forward region. Fig.\,\ref{xgamrapcuts}a shows that 
with the cuts $\eta^\gamma,\eta^{jet}>0$, the resolved photon 
component makes up for a major part of the cross section, but mainly 
consists of quarks, while for $\eta^\gamma>0.5,\,\eta^{jet}>1.5$, 
the gluon in the photon contributes about 40\% to the total cross 
section, as shown in Fig.\,\ref{xgamrapcuts}b. 

\vspace*{-0.6cm}

\clearpage

\subsection{The gluon distribution in the proton}

\begin{wrapfigure}[18]{r}[0.cm]{6cm}
\vspace*{-0.8cm}
\epsfxsize=6.cm 
\epsfbox{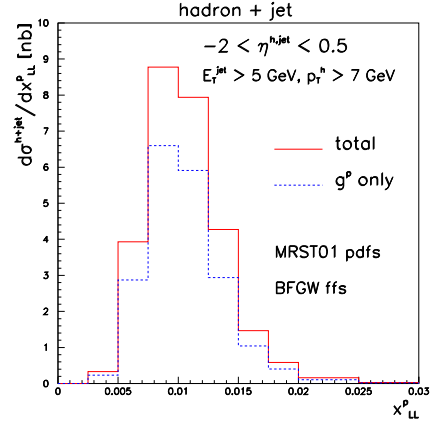}
\caption{The gluon (from the proton) 
contribution to the $h^\pm$+jet 
cross section at small rapidities. \label{hadxp}}
\end{wrapfigure}
The gluon distribution in the proton is large for small $x^p$, 
corresponding to small rapidities.
%  at fixed $p_T$ values. 
%according to eq.~(\ref{xll}). 
From Fig.\,\ref{hadxp} one can indeed see that the relative contribution 
of subprocesses initiated by a gluon from the proton to the 
$h^\pm$ + jet cross section is about 73\%   
if both rapidities are restricted to the backward region,  
$-2<\eta,\eta^{jet}<0.5$.
The analogous is {\it not} true for the prompt photon + jet cross section, 
where the 
gluon contribution makes up only 13\% 
of the total cross section if the rapidities are restricted to the range 
$-2<\eta^\gamma,\eta^{jet}<0$.
%as can be seen from Fig.~\ref{hadxp}.
This is due to the fact that at small  $x^p$ values, $x^\gamma$ is large, 
such that direct initial photons should dominate. 
However, the subprocess  
$g^{p}+\gamma\to \gamma \rm{(direct)}\,+$\,jet  
does not exist at leading order, 
and the subprocess $g^{p}+\gamma\to q\,+$\,jet, where the quark 
subsequently fragments into a photon, is suppressed by isolation. 
%\begin{wrapfigure}[13]{l}[0.cm]{7.5cm}
\begin{figure}[htb]
\begin{center}
%\vspace*{-.9cm}
\epsfxsize=7.5cm 
\epsfbox{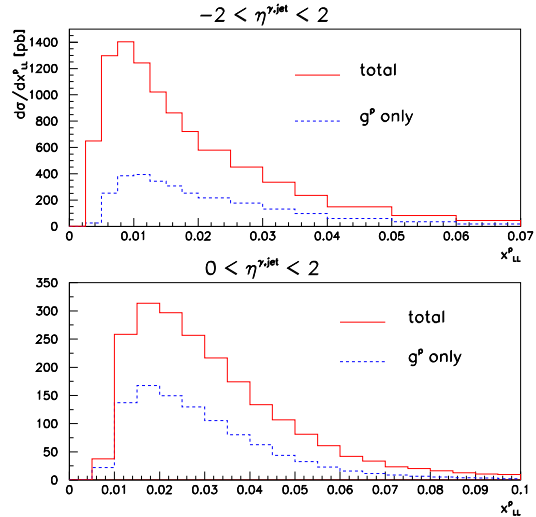}
\caption{The relative contribution of 
gluon initiated processes to the $\gamma$ + jet cross section 
can be enhanced by the rapidity cuts $0<\eta^\gamma,\eta^{jet}<2$.
The $p_T$ cuts are $E_T^{jet}> 5$\,GeV, $p_T^\gamma > 6$\,GeV.}
\label{xprapcuts}
\end{center}
\end{figure} 
%respectively small $x^p$.
Therefore, in the case of the $\gamma$ + jet cross section, the subprocess 
$g^{p}+q^{\gamma}\to \gamma+q$ is the 
dominant one involving  $g^p$ whereas in the case of the hadron+jet cross
section, $g^{p}+\gamma\to q+\bar{q}$ is dominant at small $x^p$. 
The virtue of this behaviour of the prompt photon cross section is that 
the region where $g^p$ initiated subprocesses are important 
is {\it not}  restricted to negative rapidities. 
As can be seen from Fig.\,\ref{xprapcuts}, the rapidity cuts 
$0<\eta^{\gamma,jet}<2$ enhance the relative contribution of 
the gluon $g^p$ to the $\gamma$ + jet cross section, 
because they select a region where $g^{p}+q^{\gamma}$ initiated 
subprocesses are important.
Note that this rapidity domain is 
more accessible experimentally than the very backward region.

%\clearpage

\section{Comparison to HERA data}

\subsection{Charged hadrons}

\begin{wrapfigure}[17]{r}[0.cm]{6.5cm}
\vspace*{-1.2cm}
\epsfxsize=6.5cm 
\epsfbox{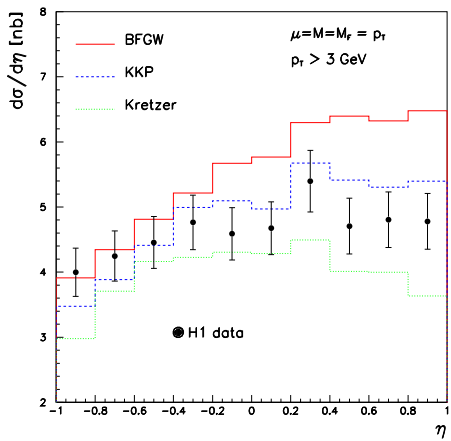}
\caption{Predictions obtained with different fragmentation 
functions 
compared to H1 data.\label{h3}}
\end{wrapfigure}
For the case of single charged hadron production, only data 
for $h^\pm$+X, but not for $h^\pm$+jet
+X are available so far\,\cite{Adloff:1998vt}. 
The analysis uses a minimum $p_T$ of 3\,GeV for the hadron, which is 
rather close to the non-perturbative regime. This induces a
large theoretical uncertainty due to differences in the parametrisation of 
the fragmentation functions\,\cite{Kretzer:2000yf,Kniehl:2000fe,Bourhis:2000gs} 
at low $p_T$ 
(see Fig.\,\ref{h3}), 
and also a large scale dependence. 
If a minimum $p_T$ of 7\,GeV is chosen for the hadron, 
the cross section is reduced, but the theoretical uncertainty 
is much smaller\,\cite{Fontannaz:2002nu}.
%, as can be seen by comparing Figs.~\ref{h3} and \ref{h7}.

\subsection{Prompt photons (inclusive)}
\begin{wrapfigure}[15]{r}[0.cm]{6.5cm}
\vspace*{-1.cm}
%\begin{figure}[htb]
\centerline{\epsfxsize=6.5cm 
\epsfbox{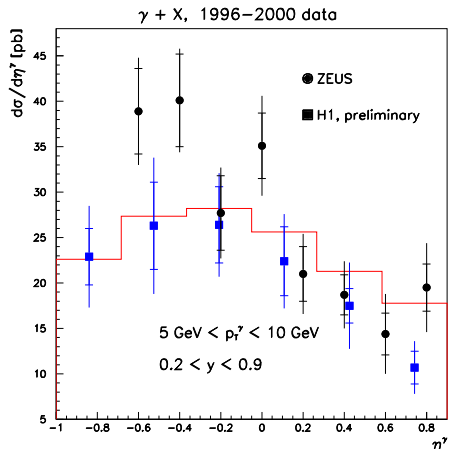}}
\vspace*{-0.7cm}
\caption{Comparison of inclusive prompt photon data
with the NLO QCD prediction.\label{gax}}
\end{wrapfigure}
For inclusive prompt photon production, ZEUS 
data\,\cite{Breitweg:1999su,Chekanov:2001aq} as well as preliminary 
H1 data\,\cite{h1} are available\footnote{In order to be able compare 
to the ZEUS data taken at $E_p=820$\,GeV and 0.2\,$<y<$\,0.9, the 
H1 data taken at $E_p=920$\,GeV and 0.2\,$<y<$\,0.7 have been corrected 
using PYTHIA\,\cite{h1}.}. 
In Fig.\,\ref{gax}, 
one observes that the ZEUS data are above the NLO prediction 
in the backward region. At large rapidities, there is  
a trend that theory is above both H1 and ZEUS data. 
As already mentioned in the introduction, this effect is very likely 
due to photon isolation: A sizeable amount of 
hadronic transverse energy 
in the 
isolation cone may stem from the underlying event.
Therefore, even
\newpage
direct photon events may be rejected, 
leading to a decrease of the cross section.
This fact cannot be simulated by a partonic  Monte Carlo, such that 
the NLO predictions tend to be above the data in kinematic regions where 
the underlying event activity is expected to be large.

\subsection{Prompt photon + jet}
\begin{wrapfigure}[22]{r}[0.cm]{7.5cm}
\vspace*{-0.8cm}
\epsfxsize=7.5cm 
\epsfbox{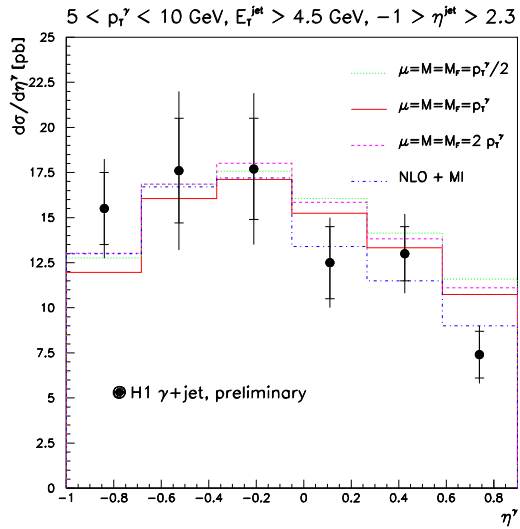}
\vspace*{-0.5cm}
\caption{Comparison of preliminary H1 data to the 
NLO prediction with different scale choices. 
NLO+MI denotes the result where the partonic 
prediction has been corrected for multiple interaction effects.\label{h1eta}}
\end{wrapfigure}
In the rapidity distribution for the $\gamma$ + jet 
cross section, the effect mentioned above is again visible at large rapidities. 
To estimate its impact, H1 recently made an analysis\,\cite{h1,Lemrani:2003mj}
where the NLO result is corrected for multiple interaction 
effects. As can be seen from Fig.\,\ref{h1eta}, 
the corrections are indeed most pronounced in the forward region where 
the underlying event activity from the resolved photon remnants
is expected to be larger, and they improve the agreement 
between data and theory.  Fig.\,\ref{h1eta} also shows that the 
$\gamma$+jet cross section is very stable with respect to scale 
variations.

\subsection{Study of intrinsic $\langle k_T\rangle$}

\begin{figure}[htb]
\centerline{
\epsfxsize=7.8cm 
\epsfbox{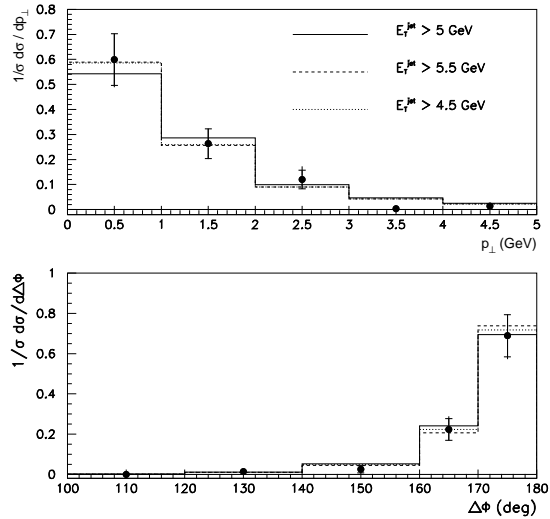}
}
\caption{ZEUS data for the normalised cross sections 
differential in $p_{\perp}$ and $\Delta\phi$ compared to the 
EPHOX prediction for different cuts on the minimum jet 
transverse energies.\label{kt}}
\end{figure}

The ZEUS collaboration made an analysis on 
prompt photon + jet data to study the 
"intrinsic" parton transverse momentum $\langle k_T\rangle$
in the proton\,\cite{Chekanov:2001aq}.
To this aim, the $\langle k_T\rangle$\,-\,sensitive observables
 $p_{\perp}$, the photon momentum component
perpendicular  to the jet direction, and 
$\Delta\phi$, the azimuthal acollinearity between  photon
and jet, have been studied.
To suppress contributions to  $\langle k_T\rangle $ from the 
 resolved photon, the cut 
$x_{\gamma}^{obs}>0.9$ has been imposed.
To minimise calibration uncertainties, only normalized cross sections
have been considered, as an additional  $\langle k_T\rangle $ mainly 
changes the {\it shape} rather than the absolute value of the cross 
section.  
The best fit with PYTHIA 6.129 lead to the result
$\langle k_T\rangle=1.69\pm 0.18({\rm stat}){+0.18\atop -0.20}$(sys)\,GeV,
which includes the shower contribution to $\langle k_T\rangle $.
On the other hand, the NLO program EPHOX is able to describe
the data without including an additional $\langle k_T\rangle $.
This is shown in Fig.~\ref{kt}, where also the minimum $E_{T}^{\rm jet}$
has been varied in order to estimate the impact of an uncertainty 
in the determination of the jet energy 
 and of choosing symmetric cuts ($p_T^\gamma>5$\,GeV has been used in 
 the experimental analysis). 
%\begin{wrapfigure}[15]{r}[0.cm]{6.5cm}
%\vspace*{-1.5cm}

\section{Conclusions}

The reactions $\gamma\,p\to \gamma$\,+\,jet\,+\,X and 
$\gamma\,p\to h^\pm$\,+\,jet\,+\,X 
offer the possibility to constrain the gluon density in the photon
and in the proton. It has been shown that appropriate 
rapidity cuts can enhance the relative contribution of  
gluon initiated processes. 

Comparing the predictions of the partonic Monte Carlo NLO program EPHOX 
to HERA data, the following observations can be made: 
For inclusive single charged hadron production, the 
H1 data are well described,  but the theoretical uncertainties 
from fragmentation functions and scale variations are large. 
These large uncertainties can mainly be 
attributed to the fact that a minimal $p_{T}$ of 3\,GeV for the hadron 
is too close to the non-perturbative regime. 
A $p_{T}^{\rm{min}}$ of 7\,GeV improves the stability of the theoretical
predictions. 

The rapidity distributions for inclusive prompt photon production 
as well as for $\gamma$+jet show an interesting feature: 
At large rapidities, the NLO prediction always tends to 
overshoot the data. It has been argued that this behaviour might be 
attributed to photon isolation: 
Due to hadronic activity stemming from the underlying event 
in the isolation cone, even direct photons in the final state 
may be rejected by the isolation cut, 
thus leading to a lower cross section 
than the (partonic) Monte Carlo prediction, especially in the 
forward region where the probability of  
resolved photon remnants is higher. 
A recent H1 analysis which corrects the NLO prediction 
bin per bin for multiple interaction effects corroborates this 
explanation. 

Finally, the predictions of EPHOX have been compared to 
a ZEUS analysis on prompt photon+jet data where the intrinsic 
$\langle k_T\rangle$ of the proton is studied, and
it has been found that the NLO prediction describes the data 
very well without introducing an extra $\langle k_T\rangle$.

\section*{Acknowledgements}
I would like to thank the organisers of the Ringberg workshop 
for an interesting and pleasant meeting, 
and J.~Gayler, R.~Lemrani and P.~Bussey for discussions on the data. 
I also wish to thank my collaborators M.~Fontannaz and J.~Ph.~Guillet.

\end{document}